\documentclass[conference]{IEEEtran}
\IEEEoverridecommandlockouts
\usepackage{geometry}
\usepackage{cite}
\usepackage{amsmath,amssymb,amsfonts}
\usepackage{algorithmic}
\usepackage{graphicx}
\usepackage{textcomp}
\usepackage{xcolor}
\usepackage{graphicx}
\usepackage[caption=false,font=footnotesize]{subfig}

\def\BibTeX{{\rm B\kern-.05em{\sc i\kern-.025em b}\kern-.08em
    T\kern-.1667em\lower.7ex\hbox{E}\kern-.125emX}}

\renewcommand{\baselinestretch}{0.996}
\begin{document}

\title{Near-Field Wideband Channel Estimation for Extremely Large-Scale RIS-Aided Communication Systems
}

\author{\IEEEauthorblockN{Lanqing Zhi, Hongwei Wang, Lingxiang Li, and Zhi Chen}
\thanks{This work was supported by the Fundamental Research Funds for the Central University under Grant ZYGX2022J029 and the Natural Science Foundation of Sichuan Province under Grant 2026NSFSC0401.}
\IEEEauthorblockA{\textit{National Key Laboratory of Wireless Communications} \\
\textit{University of Electronic Science and Technology of China}, Chengdu, P.R. China\\
hongwei\_wang@uestc.edu.cn}
}

\maketitle

\begin{abstract}
This paper studies wideband channel estimation for OFDM systems assisted by extremely large RIS (XL-RIS).  Due to the large aperture of XL-RISs, the user equipment may operate in the near-field region, while the base station–XL-RIS link remains in the far field, leading to a cascaded channel with hybrid near-field and far-field characteristics. Moreover, wideband effects further complicate channel estimation in mmWave/THz systems. To address these challenges, we propose a frequency-independent orthogonal dictionary by augmenting the discrete Fourier transform (DFT) matrix with additional parameters, which enables an efficient representation of the wideband cascaded channel using a two-dimensional block-sparse structure. Based on this property, the considered channel estimation problem is effectively solved within a tailored compressed sensing framework. Simulation results demonstrate that the proposed method significantly outperforms conventional polar-domain channel estimation approaches in 
terms of estimation accuracy.

\end{abstract}


\section{Introduction}
 
As next-generation wireless communication systems evolve rapidly, the millimeter-wave/terahertz (mmWave/THz) band has attracted considerable attention due to its abundant spectrum resources and ultra-high data rate potential~\cite{petrov2016terahertz}. It is widely regarded as a key enabler of the ever-increasing demand for ultra-high capacity and ultra-low latency by future networks~\cite{10494372}. However, mmWave/THz propagation faces several critical challenges, such as weak scattering capability, limited coverage range, and severe path loss~\cite{rappaport2019wireless}. To address these issues, intelligent reflecting surfaces (RISs) have emerged as a promising auxiliary technology. By introducing programmable phase-shifting elements, RISs can effectively reconfigure propagation channels, thereby enhancing coverage performance and improving signal quality~\cite{hu2018beyond}. In particular, RISs can create virtual line-of-sight (LoS) links for UEs, and thus mitigate coverage shortages in mmWave/THz communications~\cite{su2023wideband}. 

Accurate channel state information (CSI) is essential for an RIS to align passive reflections and deliver the promised beamforming gains. However, RIS elements are passive and unable to process signals, making channel estimation in RIS-assisted systems significantly more challenging. The least-squares (LS) method can be adopted to estimate CSI in RIS-assisted systems. However, its high training overhead limits practical deployment. In the mmWave/THz bands, free-space loss and molecular absorption suppress multipath components, yielding sparse channels~\cite{bhardwaj2024generalized}. Exploiting this sparsity property, several compressive sensing (CS) algorithms such as the orthogonal matching pursuit (OMP) and the sparse Bayesian learning (SBL) are employed to acquire CSI while significantly reducing training overhead~\cite{10192507,9801875,9103231}. Off-grid CS methods such as atomic-norm minimization~\cite{he2021channel} and high-order tensor factorizations~\cite{10552118} are introduced to further improve estimation accuracy by alleviating grid mismatch.

As the RIS aperture and operating frequency increase, the Rayleigh distance stretches to tens or even hundreds of meters, making UEs more likely to be located in the near-field region~\cite{pan2023ris,10758747}. This near-field characteristic complicates the acquisition of CSI, making channel estimation highly challenging in RIS-assisted mmWave/THz communication systems~\cite{11310291}. Directly mapping physical near-field channels onto the conventional spatial Fourier basis induces pronounced energy spreading. Therefore, the angular sparsity widely exploited for far-field channel collapses, and the conventional far-field channel estimation approaches are no longer applicable. To address this, several approaches have been introduced to estimate the near-field channels, e.g., the polar-domain-based~\cite{lei2023channel} and the block-sparsity-based solutions~\cite{10437620,11316669}. 

On the other hand, orthogonal frequency-division multiplexing (OFDM) has been widely used in mmWave/THz systems to improve spectral efficiency, since mmWave/THz bands provide extremely abundant spectrum resources~\cite{zheng2020fast}. Exploiting the strong inter-subcarrier channel correlation, joint CSI estimation across subcarriers outperforms per-subcarrier processing pattern~\cite{wang2025compressive}. It is commonly assumed that channels across subcarriers share the same sparse pattern. Consequently, algorithms such as simultaneous orthogonal matching pursuit (SOMP)~\cite{cui2021near,cui2022channel} can be leveraged to estimate OFDM channels. In wideband systems, however, the wideband effect cannot be ignored, especially in mmWave/THz systems whose bandwidth is exceptionally large.

In this work, we address the channel estimation problem for RIS-assisted OFDM systems under both near-field conditions and the wideband effect. Specifically, we propose a unified modified DFT dictionary that is valid across all subcarriers, with which the cascaded channel exhibits a two-dimensional block-sparse structure. Consequently, near-field cascaded channel estimation problem can be efficiently performed by a two-dimensional pattern-coupled sparse Bayesian learning (2D-PCSBL) algorithm~\cite{fang2016two}. Numerical results show that the proposed method achieves significantly higher estimation accuracy than the conventional polar-domain-based approach.
 
\section{System Model and Signal Model}
\subsection{System Model and Notations}

We consider the downlink channel estimation problem for a mmWave/THz OFDM system in which a XL-RIS is deployed to assist a base station (BS) in serving multiple single-antenna users. The BS is equipped with a uniform linear array (ULA) consisting of $N_t$ antennas, while the XL-RIS is a uniform planar array (UPA) of $N = N_y \times N_z$ passive reflecting elements. The system operates at a carrier frequency $f_c$ with wavelength $\lambda_c = c/f_c$, where $c$ denotes the speed of light, and the inter-element spacing is $d = \lambda_c / 2$. The total bandwidth is $B$, and $P$-point OFDM is adopted. The frequency of the $p$-th subcarrier is given by $f_p = f_c + \frac{(2p-P)B}{2P}$ with $\lambda_p = {c}/{f_p}$.

During the downlink training phase, the BS transmits pilot sequences for channel estimation. Each user independently receives the training signals and estimates its own channel. Therefore, we focus on the channel estimation for a specific user. We denote $\boldsymbol{G}_p \in \mathbb{C}^{N \times N_t}$ as the BS-to-RIS channel at the $p$-th subcarrier, and $\boldsymbol{h}_p \in \mathbb{C}^{N \times 1}$ as the channel from the XL-RIS to the UE. Each XL-RIS element imposes a phase shift to the incident signal. The corresponding reflection coefficients at the $t$-th time instant are represented as
\begin{align}
\boldsymbol{s}(t) = \left[ e^{j\alpha_1(t)}, \dots, e^{j\alpha_N(t)} \right]^T \in \mathbb{C}^{N\times 1},
\end{align}
where $\boldsymbol{s}(t)$ denotes the reflection coefficient. The signal received by the user at the $t$-th time instant on subcarrier $p$ is
\begin{align}
    y_p(t) &= \boldsymbol{h}_p^H \mathrm{diag}(\boldsymbol{s}(t)) \boldsymbol{G}_p \boldsymbol{f}(t) x_p(t) + n_p(t) \notag\\
    &=\boldsymbol{s}(t)^T \mathrm{diag}(\boldsymbol{h}_p^H) \boldsymbol{G}_p \boldsymbol{f}(t) x_p(t) + n_p(t) \notag\\
    &= \boldsymbol{s}(t)^T \boldsymbol{H}_p \boldsymbol{f}(t) x_p(t) + n_p(t) \notag\\
    &= \left( \boldsymbol{f}^T(t) \otimes \boldsymbol{s}^T(t) \right) \text{vec}(\boldsymbol{H}_p)x_p(t) + n_p(t),
\end{align}
where $\boldsymbol{f}(t) \in \mathbb{C}^{N_t\times 1}$ is the beamforming vector, $n_p(t) \sim \mathbb{CN}(0,\sigma_n^2)$ is additive white Gaussian noise, $x_p(t)$ is the transmitted symbol (for simplicity, $x_p(t)$ is set to $1$ during channel estimation stage), and $\boldsymbol{H}_p \triangleq \mathrm{diag}(\boldsymbol{h}_p^H) \boldsymbol{G}_p \in \mathbb{C}^{N \times N_t}$ is the cascaded BS-RIS-UE channel for the $p$-th subcarrier. Collecting $T$ snapshots, the received signal on the $p$-th subcarrier is 
\begin{align}
\boldsymbol{y}_p = \boldsymbol{C}^T \text{vec}(\boldsymbol{H}_p)+ \boldsymbol{n}_p,
\end{align}
where $\boldsymbol{C} \triangleq  [ \cdots, ( \boldsymbol{f}^T(t) \otimes \boldsymbol{s}^T(t))^T, \cdots  ] \in \mathbb{C}^{NN_t \times T}$ and $\boldsymbol{n}_p\triangleq [n_p(1),\cdots,n_p(T)]^T$. By stacking the received vectors across subcarriers, we obtain
\begin{align}
\boldsymbol{Y} = \boldsymbol{C}^T \boldsymbol{H} + \boldsymbol{N},
\label{Y}
\end{align}
where $\boldsymbol{Y} \triangleq [\boldsymbol{y}_1 \cdots \boldsymbol{y}_P]$, $\boldsymbol{H} \triangleq [\text{vec}(\boldsymbol{H}_1)\ \cdots\ \text{vec}(\boldsymbol{H}_P)]$ and 
$\boldsymbol{N} \triangleq [\boldsymbol{n}_1 \cdots \boldsymbol{n}_P]$.

\subsection{Wideband Channel Model}

In this work, we follow the deployment guidelines widely adopted in the literature, where the RIS is deployed on the user side~\cite{large2023multi}. In this case, $\boldsymbol{G}_p$ is approximated by a far-field channel model, while $\boldsymbol{h}_p$ is characterized using a near-field channel model. Specifically, $\boldsymbol{G}_p$ is expressed as
\begin{align}
\boldsymbol{G}_p = \sqrt{\tfrac{N_t N}{L}} \sum_{l=1}^L \rho_{l,p} \, \boldsymbol{b}(\theta_{l,1},\theta_{l,2},f_p) \, \boldsymbol{a}^H(\vartheta_l,f_p),
\end{align}

where $\rho_{l,p}$ is the complex gain of the $l$-th path; $\vartheta_l$ is its BS AoD; $\{\theta_{l,1},\,\theta_{l,2}\}$ are the RIS elevation/azimuth angles; $\boldsymbol{a}(\cdot,\cdot)$ denotes the BS far-field steering vectors; and $\boldsymbol{b}(\cdot,\cdot,\cdot)$ is the RIS far-field steering vector. $\boldsymbol{a}(\cdot,\cdot)$ is expressed as
\begin{align}
\boldsymbol{a}(\vartheta,f_p) \triangleq \tfrac{1}{\sqrt{N_t}} \left[\cdots, e^{jk_p n_t d \sin\vartheta}, \cdots \right]^T
\end{align}
where $k_p\triangleq 2\pi/\lambda_p$, and $n_t \in \{0,1,\ldots,N_t-1\}$ denote the BS antenna index. $\boldsymbol{b}(\cdot,\cdot,\cdot)$ is given by
\begin{align}
\boldsymbol{b}(\theta_1,\theta_2,f_p) = \tfrac{1}{\sqrt{N}} \left[ \cdots e^{jk_p d (n_z \Psi_a + n_y \Psi_e)} \cdots \right]^T
\label{FF_SV}
\end{align}
where $\Psi_a(\theta_1) \triangleq \sin \theta_1,\Psi_e(\theta_1,\theta_2) \triangleq \cos \theta_1 \sin \theta_2$, $n_y \in \{0,\ldots,N_y-1\}$ and $n_z \in \{0,\ldots,N_z-1\}$ index the RIS elements along the $y$- and $z$-axes, respectively.

Similarly, $\boldsymbol{h}_p$ is expressed as
\begin{align}
\boldsymbol{h}_p &= \sqrt{\tfrac{N}{K}} \sum_{k=1}^K \xi_{k,p} e^{-j\frac{2\pi}{\lambda_p} r_k} \boldsymbol{c}(\varphi_{k,1},\varphi_{k,2}, f_p, r_k) \notag \\
&= \sqrt{\tfrac{N}{K}} \sum_{k=1}^K \tilde{\xi}_{k,p} \, \boldsymbol{c}(\varphi_{k,1},\varphi_{k,2}, f_p, r_k)
\label{h}
\end{align}
where $\tilde{\xi}_{k,p} \triangleq \xi_{k,p} e^{-j2\pi f_p \tau_k}$; $K$ is the path number between the XL-RIS and the UE; $k$ denotes the path index; $\xi_{k,p}$ is the complex path gain; $\varphi_{k,1}$ and $\varphi_{k,2}$ denote the azimuth and elevation angles of departure from the RIS; $\boldsymbol{c}(\cdot)$ denotes the near-field steering vector, which is given by
\begin{align}
\boldsymbol{c}(\varphi_1,\varphi_2,f_p,r) \triangleq \tfrac{1}{\sqrt{N}} \left[ \cdots, e^{jk_p(r_{n_y n_z} - r)}, \cdots \right]^T,  
\end{align}
where $r_{n_y n_z}$ is the distance between the UE/scatterer and the $(n_y,n_z)$-th RIS element. According to the geometric relationship, we have
\begin{align}
r_{n_y n_z} = \sqrt{r^2 + d^2(n_y^2 + n_z^2) + 2r d (n_z \Psi_a + n_y \Psi_e)}
\label{true_distance}
\end{align}
where $r$ is the distance between the reference element of the XL-RIS to the user/scatterer, $\Psi_a = \sin\varphi_1$, and $\Psi_e = \cos\varphi_1\sin\varphi_2$.

Directly using~\eqref{true_distance} complicates the near-field channel estimation. To deal with this problem, we apply the following approximation~\cite{cui2022channel}
\begin{align}
r_{n_y n_z}& \approx r + d(n_z \Psi_a + n_y \Psi_e) \notag\\
 & \qquad \qquad + \frac{d^2 \left[ n_z^2 (1 - \Psi_a^2) + n_y^2 (1 - \Psi_e^2) \right]}{2r}
\label{r_app}
\end{align}
Based on this approximation, the near-field steering vector can be composed as
\begin{align}
    \boldsymbol{c}(\varphi_1,\varphi_2, f_p,r) \approx \boldsymbol{b}(\varphi_1, \varphi_2,f_p) \circ \boldsymbol{d}(r,\varphi_1,\varphi_2,f_p)
    \label{c}
\end{align}
where $\circ$ denotes the Hadamard product, $\boldsymbol{b}(\cdot)$ is the far-field response in~\eqref{FF_SV}, and $\boldsymbol{d}(\cdot)$ is given by

\begin{align}
\boldsymbol{d}(r,\varphi_1,\varphi_2,f_p)
&= \left[ \cdots, e^{jk_p\frac{d^2}{2 \mu(r,\varphi_1,\varphi_2)}} , \cdots \right]^T \notag \\
& \triangleq \boldsymbol{d}(\mu(r,\varphi_1,\varphi_2),f_p)
\label{d}
\end{align}
where $\mu(r,\varphi_1,\varphi_2) = \frac{r}{n_z^2(1-\Psi_a^2)+n_y^2(1-\Psi_e^2)}$.

\subsection{Problem Formulation}
Following the received model in~\eqref{Y}, our objective is to estimate the channel $\boldsymbol{H}$ from the received observations $\boldsymbol{Y}$. However, $\boldsymbol{H}$ is a high dimensional matrix, which poses a significant challenge for channel estimation. To reduce pilot overhead and mitigate the joint effects of spherical-wave propagation and beam squint in near-field wideband scenarios, we propose a novel channel representation matrix that exhibits a 2D block-sparse structure. Based on this representation, the channel $\boldsymbol{H}$ can be efficiently and accurately estimated by the proposed algorithm. 

\section{Proposed Method}
\subsection{Block Sparsity for a Specific Sub-Carrier Channel}

For a specific subcarrier, the far-field channel $\boldsymbol{G}_p$ can be sparsely represented by
\begin{equation}
    \boldsymbol{G}_p = \boldsymbol{U}_p \boldsymbol{\Lambda}_p \boldsymbol{V}_p^H
    \label{G_p}
\end{equation}
where $\boldsymbol{U}_p \in \mathbb{C}^{N \times N}$ and $\boldsymbol{V}_p \in \mathbb{C}^{N_t \times N_t}$ are the 2D-DFT matrix and the DFT matrix, respectively, while $\boldsymbol{\Lambda}_p \in \mathbb{C}^{N \times N_t}$ is a sparse matrix with only $L$ non-zero elements. For $\boldsymbol{h}_p$,  however, this DFT-based sparse representation method is not suitable because $\boldsymbol{h}_p$ is in general no longer sparse in the angular domain. To this end, in our previous work~\cite{10758747}, we devised a specific dictionary $\boldsymbol{D}_p$ and $\boldsymbol{h}_p$ is block-sparsely represented by
\begin{align}
    \boldsymbol{h}_p = \boldsymbol{D}_{p}  \boldsymbol{\beta}_{p}
\label{h_p}
\end{align}
where $\boldsymbol{\beta}_p$ is a block-sparse vector with its sparsity level being on the order of ${1}/{\sqrt{N}}$. Essentially, $\boldsymbol{D}_p$ is a modified version of the DFT matrix, which is given by
\begin{align}
    \boldsymbol{D}_{p} = \text{diag}(\boldsymbol{d}(\mu, f_p)) \boldsymbol{U}_p
    \label{D_p}
\end{align}
where $\boldsymbol{U}_p$ is the 2D-DFT matrix. Apparently, $\boldsymbol{D}_{p}$ is a unitary matrix.

Based on the sparse representation of $\boldsymbol{G}_p$ and $\boldsymbol{h}_p$, the cascaded channel $\boldsymbol{H}_p$ can be written as
\begin{equation}
\begin{aligned}
\boldsymbol{H}_p &= \text{diag}(\boldsymbol{h}_p^H) \boldsymbol{G}_p \stackrel{(a)}{=} \boldsymbol{h}_p^* \bullet \boldsymbol{G}_p \\
&\stackrel{(b)}{=} (\boldsymbol{D}_{p}^* \boldsymbol{\beta}_{p}^*) \bullet (\boldsymbol{U}_p \boldsymbol{\Lambda}_p \boldsymbol{V}_p^H) \\
&\stackrel{(c)}{=} (\boldsymbol{D}_{p}^*\bullet \boldsymbol{U}_p) (\boldsymbol{\beta}_{p}^* \otimes (\boldsymbol{\Lambda}_p \boldsymbol{V}_p^H)) \\
&\stackrel{(d)}{=} ((\text{diag}(\boldsymbol{d}^*(\mu,f_p)) \boldsymbol{U}_p^*) \bullet\boldsymbol{U}_p) (\boldsymbol{\beta}_p^* \otimes \boldsymbol{\Lambda}_p) \boldsymbol{V}_p^H \\
&\stackrel{(e)}{=} \text{diag}(\boldsymbol{d}^*(\mu,f_p)) (\boldsymbol{U}_p^*\bullet \boldsymbol{U}_p) (\boldsymbol{\beta}_p^* \otimes \boldsymbol{\Lambda}_p) \boldsymbol{V}_p^H \\
&\stackrel{(f)}{=} \boldsymbol{\Theta}_p \boldsymbol{\Phi}_p \boldsymbol{V}_p^H
\end{aligned}
\end{equation}
where $\bullet$ in (a) denotes the transposed Khatri-Rao product; (b) is from~\eqref{G_p} and~\eqref{h_p}; in (c) and (e) we apply the property of the transposed Khatri-Rao product; (d) comes from~\eqref{D_p}; and in (f), we define $\boldsymbol{\Theta}_p = \text{diag}(\boldsymbol{d}^*(\mu,f_p)) (\boldsymbol{U}_p^*\bullet \boldsymbol{U}_p) \in \mathbb{C}^{N \times N^2}$ and $\boldsymbol{\Phi}_p=\boldsymbol{\beta}_p^* \otimes \boldsymbol{\Lambda}_p \in \mathbb{C}^{N^2 \times N_t}$.

Since $\boldsymbol{U}_p$ is a 2D-DFT matrix, the columns of $\boldsymbol{U}_p^{*}\bullet\boldsymbol{U}_p$ differ from one another only by circular shifts in the indices; meanwhile, $\mathrm{diag}(\boldsymbol{d}^{*}(\mu,f_p))$ merely scales the rows and does not introduce mixing across columns, and therefore does not alter the equivalence relation among these columns. Consequently, the $N^{2}$ columns of $\boldsymbol{\Theta}_p$ can be grouped into $N$ equivalence classes. For brevity and without loss of generality, we select one column from each equivalence class as a representative, specifically taking the  $N$ columns, to obtain
\begin{equation}
\boldsymbol{\Theta}_{N,p} \triangleq \boldsymbol{\Theta}_p(:, 1:N) \in \mathbb{C}^{N \times N}
\end{equation}
We then have
\begin{equation}
\boldsymbol{\Theta}_p = \boldsymbol{\Theta}_{N,p} \mathcal{P}
\end{equation}
where $\mathcal{P} \in \{0,1\}^{N \times N^2}$ is an aggregation matrix, which serves to merge columns belonging to the same difference index. Thus, $\mathcal{P}$ consists of $N$ permutation blocks of size $N\times N$.

Correspondingly, the same aggregation operation is applied to the matrix $\boldsymbol{\Phi}_p$
\begin{equation}
\boldsymbol{\Phi}_{N,p} \triangleq \mathcal{P} \boldsymbol{\Phi}_p
\end{equation}
The channel $\boldsymbol{H}_p$ can then be written as
\begin{equation}
\boldsymbol{H}_p = \boldsymbol{\Theta}_{N,p} \boldsymbol{\Phi}_{N,p} \boldsymbol{V}_p^H 
\end{equation}

To facilitate sparse estimation, we vectorize $\boldsymbol{H}_p$.
Using $\mathrm{vec}(\boldsymbol{A}\boldsymbol{X}\boldsymbol{B})=(\boldsymbol{B}^T\otimes\boldsymbol{A})\mathrm{vec}(\boldsymbol{X})$, we obtain
\begin{equation}
\begin{aligned}
    \mathop{\mathrm{vec}}(\boldsymbol{H}_p) &= (\boldsymbol{V}_p^* \otimes \boldsymbol{\Theta}_{N,p}) \mathop{\mathrm{vec}}(\boldsymbol{\Phi}_{N,p})\\
    &=(\boldsymbol{V}_p^* \otimes \boldsymbol{\Theta}_{N,p}) \mathop{\mathrm{vec}}(\mathcal{P} (\boldsymbol{\beta}_p^* \otimes \boldsymbol{\Lambda}_p))\\
    &=(\boldsymbol{V}_p^* \otimes \boldsymbol{\Theta}_{N,p})(\boldsymbol{I}_{N_t}\otimes\mathcal{P}) \mathop{\mathrm{vec}} (\boldsymbol{\beta}_p^* \otimes \boldsymbol{\Lambda}_p)\\
    &=(\boldsymbol{V}_p^* \otimes \boldsymbol{\Theta}_{N,p}) \boldsymbol{x}_p
\end{aligned}
\end{equation}
where $ \boldsymbol{x}_p \triangleq (\boldsymbol{I}_{Nt}\otimes\mathcal{P}) \mathop{\mathrm{vec}} (\boldsymbol{\beta}_p^* \otimes \boldsymbol{\Lambda}_p)$.

\begin{figure}[!htp]
	\centering
	\subfloat{\includegraphics[width=0.35\textwidth]{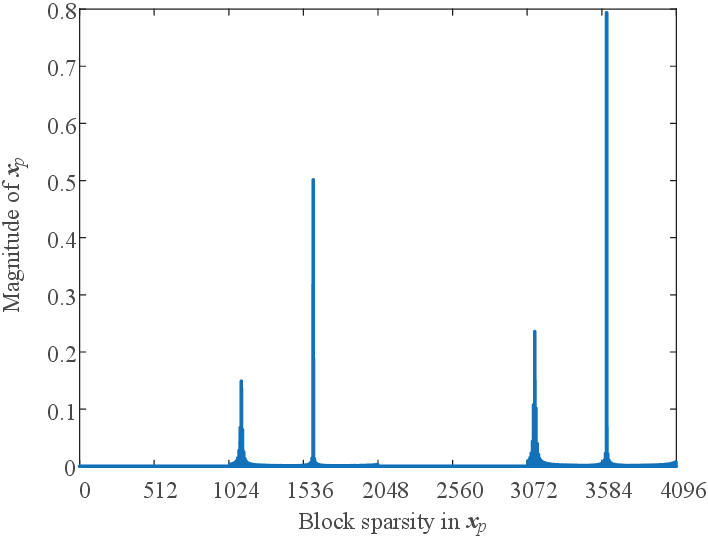}} 
	\hfil
	\caption{Block sparsity in $\boldsymbol{x}_p$.}
	\label{fig:X}
\end{figure}

Since $\boldsymbol{\beta}_p$ is a block-sparse vector, and $\boldsymbol{\Lambda}_p$ is a sparse matrix, 
it can be concluded that $\boldsymbol{\Phi}_p$ (which is the Kronecker product of 
$\boldsymbol{\beta}^{*}_{p}$ and $\boldsymbol{\Lambda}_p$) exhibits block-sparsity 
in the column dimension. In addition, because $\boldsymbol{x}_p$ consists of $N$ permutation blocks of size $N\times N$ and acts from the left, it aggregates only along the row dimension and does not disrupt the original column-wise block-sparse pattern. In other words, $\boldsymbol{x}_p$ remains block-sparse along the column dimension. A numerical simulation is implemented to verify this, in which the same setup is used as in the simulation section. Fig.~\ref{fig:X} shows that the nonzero entries are still concentrated in a few adjacent column blocks, confirming the block sparsity of $\boldsymbol{x}_p$.
\subsection{2D Block-sparse Structure}
To explore the joint sparse characteristics of wideband channels, we first construct a unified transform dictionary for all subcarriers:
$\boldsymbol{E}_{\mu} = \sqrt{N}\boldsymbol{V}_c^* \otimes \boldsymbol{\Theta}_{N,c}$,
which is parameterized by the center frequency $f_c$. Since each subcarrier operates at a different frequency $f_p$, the channel corresponding to each frequency will exhibit different sparse structures in this dictionary. Specifically, the channel on the $p$-th subcarrier can be represented as
\begin{equation}
\mathop{\mathrm{vec}}(\boldsymbol{H}_p) = \tfrac{1}{\sqrt{N}}\boldsymbol{E}_{\mu} \tilde{\boldsymbol{x}}_p
\label{H_p}
\end{equation}
where the block sparse vector $\tilde{\boldsymbol{x}}_p$ is
\begin{align}
    \tilde{\boldsymbol{x}}_p &= \sqrt{N}\boldsymbol{E}_{\mu}^H \mathop{\mathrm{vec}}(\boldsymbol{H}_p)\notag\\
    &= \sqrt{N}\boldsymbol{E}_{\mu}^H\mathop{\mathrm{vec}}(\boldsymbol{h}_p^* \bullet \boldsymbol{G}_p)\notag\\
    &= C\boldsymbol{E}_{\mu}^H\mathop{\mathrm{vec}} 
        \Bigl( (\sum\nolimits_{k=1}^K \tilde{\xi}_{k,p}^* \, \boldsymbol{c}^*(\varphi_{k,1},\varphi_{k,2}, f_p, r_k)) \notag\\
    &\qquad \bullet (\sum\nolimits_{l=1}^L \rho_{l,p} \, \boldsymbol{b}(\theta_{l,1},\theta_{l,2},f_p) \, \boldsymbol{a}^H(\vartheta_l,f_p)) \Bigr)
    \label{x}
\end{align}
where $\tilde{\xi}_{k,p} \triangleq \xi_{k,p} e^{-j2\pi f_p \tau_k}$, $C=N\sqrt{(N_tN)/(KL)}$, and
$\boldsymbol{c}(\varphi_{k,1},\varphi_{k,2}, f_p,r_k) \approx \boldsymbol{b}(\varphi_{k,1},\varphi_{k,2},f_p) \circ \boldsymbol{d}(r_k,\varphi_{k,1},\varphi_{k,2},f_p)$.

Using the previous equation~\eqref{c}, the near-field steering vector $\boldsymbol{c}(\varphi_{k,1},\varphi_{k,2}, f_p,r_k)$ can be decomposed as the Hadamard product of the far-field steering vector $\boldsymbol{b}(\varphi_{k,1},\varphi_{k,2},f_p)$ and the distance-related unit-norm vector $\boldsymbol{d}(\mu(r_k,\varphi_{k,1},\varphi_{k,2}),f_p)$. We define $\eta_p \triangleq {k_p}/{k_c}$, and $\boldsymbol{b}(\varphi_{k,1},\varphi_{k,2},f_p)$ and $\boldsymbol{d}(\mu(r_k,\varphi_{k,1},\varphi_{k,2}),f_p)$ are, respectively, equivalently formulated as
\begin{align}
    \boldsymbol{b}(\varphi_{k,1},\varphi_{k,2},f_p) 
        &=\tfrac{1}{\sqrt{N}} \left[ \cdots e^{j k_c d \eta_p (n_z \Psi_a + n_y \Psi_e)} \cdots \right]^T \notag \\
        &=\tfrac{1}{\sqrt{N}} \left[ \cdots e^{j k_c d (n_z \tilde{\Psi}_{a,p} + n_y \tilde{\Psi}_{e,p})} \cdots \right]^T \notag \\
        &=\boldsymbol{b}(\tilde{\varphi}_{k,p,1},\tilde{\varphi}_{k,p,2}, f_c)
        \label{b_new}
\end{align}
\begin{align}
\boldsymbol{d}(\mu(r_k,\varphi_{k,1},\varphi_{k,2}),f_p) 
&= \left[ \cdots, e^{jk_c \eta_p \frac{d^2}{2 \mu(r_k,\varphi_{k,p,1},\varphi_{k,p,2})}} , \cdots \right]^T \notag \\
&= \left[ \cdots, e^{jk_c \frac{d^2}{2 \mu(\tilde{r}_{k,p}, \tilde{\varphi}_{k,p,1}, \tilde{\varphi}_{k,p,2})}} , \cdots \right]^T \notag \\
&=\boldsymbol{d}(\mu(\tilde{r}_{k,p}, \tilde{\varphi}_{k,p,1}, \tilde{\varphi}_{k,p,2}), f_c)
\label{d_new}
\end{align}
\begin{align}
\boldsymbol{a}(\vartheta_l,f_p)
&= \tfrac{1}{\sqrt{N_t}} \left[\cdots, e^{jk_c\eta_p n_t d \sin\vartheta_l}, \cdots \right]^T \notag \\
&=\tfrac{1}{\sqrt{N_t}} \left[\cdots, e^{jk_c n_t d \sin\tilde{\vartheta}_{l,p}}, \cdots \right]^T \notag \\
&=\boldsymbol{a}(\tilde{\vartheta}_{l,p},f_c)
\label{a_new}
\end{align}
where
\begin{align}
    \tilde{\Psi}_{a,p}&= \eta_p \Psi_{a}, \ \tilde{\Psi}_{e,p}=\eta_p\Psi_{e} \notag  \\
\sin\tilde{\vartheta}_{l,p} & =\eta_p \sin\vartheta_{l}\notag \\
\tilde{r}_{k,p}&=\frac{n_z^2(1-\eta_p^2\Psi_{a}^2)+n_y^2(1-\eta_p^2\Psi_{e}^2)}{\eta_p[n_z^2(1-\Psi_{a}^2)+n_y^2(1-\Psi_{e}^2)]}r_k \notag \\
\mu(r_k,\varphi_{k,1},\varphi_{k,2})&= \eta_p \mu(\tilde{r}_{k,p}, \tilde{\varphi}_{k,p,1}, \tilde{\varphi}_{k,p,2})
\label{map}
\end{align}

Substituting equations~\eqref{b_new},~\eqref{d_new} and~\eqref{a_new} into equation~\eqref{x} yields:
\begin{align}
    \tilde{\boldsymbol{x}}_p 
    &= C\boldsymbol{E}_{\mu}^H\mathop{\mathrm{vec}} 
        \Bigl( (\sum\nolimits_{k=1}^K \tilde{\xi}_{k,p}^* \, \boldsymbol{c}^*(\tilde{\varphi}_{k,p,1},\tilde{\varphi}_{k,p,2}, f_c, \tilde{r}_{k,p})) \notag \\
    &\qquad \bullet (\sum\nolimits_{l=1}^L \rho_{l,p} \, \boldsymbol{b}(\tilde{\theta}_{l,p,1},\tilde{\theta}_{l,p,2},f_c) \, \boldsymbol{a}^H(\tilde{\vartheta}_{l,p},f_c)) \Bigr)
    \label{x_new}
\end{align}
From~\eqref{x_new}, the steering vectors $\boldsymbol{a}(\vartheta_l,f_p)$, $\boldsymbol{b}(\theta_{l,1},\theta_{l,2},f_p)$, $\boldsymbol{c}(\varphi_{k,1},\varphi_{k,2}, f_p, r_k)$ defined at subcarrier $f_p$ can be frequency-normalized to a common reference at $f_c$. Under this normalization, the spatial-angle parameters $\Psi_{a}$, $\Psi_{e}$, $\sin\vartheta_{l}$ and the relative effective distance $\mu(r_k,\varphi_{k,1},\varphi_{k,2})$ are mapped to $\tilde{\Psi}_{a,p}$, $\tilde{\Psi}_{e,p}$, $\sin\tilde{\vartheta}_{l,p}$ and $\mu(\tilde{r}_{k,p}, \tilde{\varphi}_{k,p,1}, \tilde{\varphi}_{k,p,2})$, respectively.

The resulting coefficient vector $\tilde{\boldsymbol{x}}_p$ remains block-sparse with respect to the transformed virtual scatterer locations. In addition,~\eqref{map} shows that the post-transformation spatial angle and range are affinely related to their pre-transformation counterparts. Hence, the block support of a near-field wideband channel drifts almost linearly with frequency, yielding strong coupling between adjacent subcarriers.

Let $\tilde{\boldsymbol{x}}_p$ denote the block-sparse coefficients at subcarrier $p$. By stacking $\tilde{\boldsymbol{x}}_p$ along the frequency dimension into a matrix, we obtain a two-dimensional block-sparse representation of the channel over an enhanced DFT dictionary $\mathbf{E}_\mu$. We refer to this property as two-dimensional support-coupled block sparsity.

Furthermore, stacking the block-sparse coefficient vectors across all subcarriers along the frequency dimension, we define
\begin{equation}
\tilde{\boldsymbol{X}} \triangleq [\tilde{\boldsymbol{x}}_1\ \cdots\ \tilde{\boldsymbol{x}}_P].
\end{equation}
Since~\eqref{H_p} holds for all subcarriers $p$, the wideband channel admits the following 2D block-sparse representation:
\begin{equation}
\boldsymbol{H} = \tfrac{1}{\sqrt{N}}\boldsymbol{E}_{\mu}\tilde{\boldsymbol{X}}.
\label{HH}
\end{equation}
Here, the nonzero entries of $\tilde{\boldsymbol{X}}$ form clustered blocks jointly over the coefficient index and the subcarrier index, which characterizes a 2D support-coupled block-sparse structure.

\begin{figure}[!htp]
	\centering
	\subfloat{\includegraphics[width=0.35\textwidth]{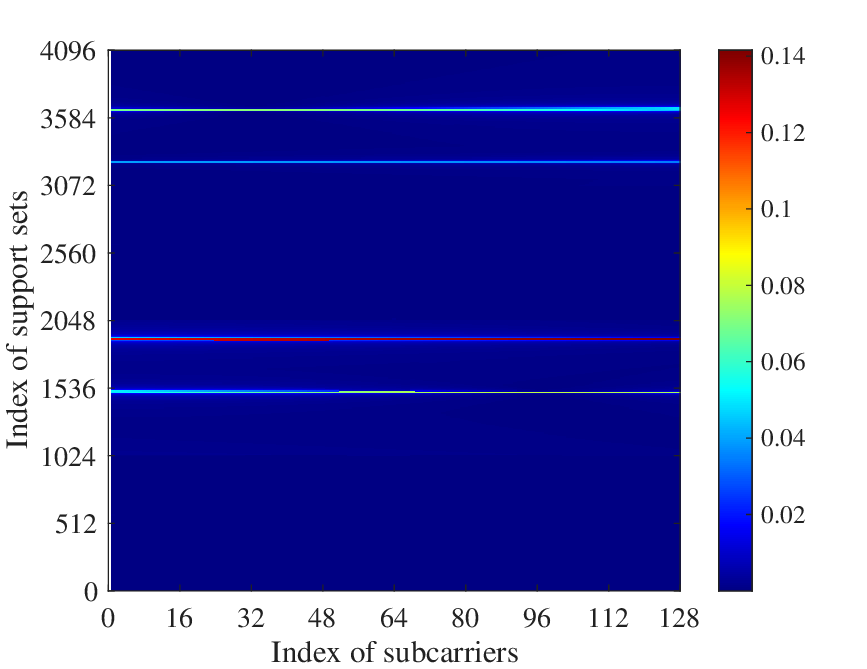}}
	\hfil
	\caption{The sparsity of the wideband channel $\boldsymbol{H}$}
	\label{fig:channel_sparsity}
\end{figure}

To validate the sparsity characteristics discussed earlier, Fig.~\ref{fig:channel_sparsity} illustrates the sparse structure of the wideband channel when adopting a commonly used modified discrete Fourier transform (DFT) dictionary. The simulation parameters are consistent with those used in simulation section. As shown in the figure, each column contains four non-zero blocks corresponding to the four propagation paths. Furthermore, the blocks across different subcarriers exhibit strong correlations, verifying the inherent sparsity of the wideband channel.

\subsection{2D Block-Sparsity-Aware Wideband Near-field Channel Estimation}
Substituting~\eqref{HH} into~\eqref{Y}, the wideband observation model can be rewritten as
\begin{equation}
\boldsymbol{Y}
=\boldsymbol{C}^T\Bigl(\tfrac{1}{\sqrt{N}}\boldsymbol{E}_{\mu}\tilde{\boldsymbol{X}}\Bigr)+\boldsymbol{N}
=\boldsymbol{\Omega}\tilde{\boldsymbol{X}}+\boldsymbol{N},
\label{YY}
\end{equation}
where the equivalent sensing matrix is defined as
$\boldsymbol{\Omega}\triangleq \tfrac{1}{\sqrt{N}}\boldsymbol{C}^T\boldsymbol{E}_{\mu}$.
Therefore, the wideband channel estimation problem is converted into recovering the 2D support-coupled block-sparse matrix $\tilde{\boldsymbol{X}}$ from the noisy measurements $\boldsymbol{Y}$.

To fully exploit both intra-subcarrier block sparsity and inter-subcarrier support coupling, we adopt a 2D pattern-coupled sparse Bayesian learning (2D-PCSBL)~\cite{fang2016two} framework to estimate $\tilde{\boldsymbol{X}}$ from~\eqref{YY}.
Specifically, 2D-PCSBL introduces neighborhood-coupled prior hyperparameters along both the coefficient index and the subcarrier index, thereby jointly leveraging the 2D block sparsity and the cross-subcarrier correlation to recover $\tilde{\boldsymbol{X}}$.

Note that, besides the adopted 2D-PCSBL framework, the compressive sensing problem in~\eqref{YY} can also be solved by other algorithms, such as BOMP~\cite{eldar2010block} or PCSBL~\cite{fang2014pattern}.
However, these methods cannot effectively exploit the 2D support-coupled block-sparse structure in $\tilde{\boldsymbol{X}}$.

\section{Numerical Simulation}
In this section, we validate the effectiveness of the proposed method via numerical simulations. The baselines include subcarrier-to-subcarrier orthogonal matching pursuit in the polar domain (P-OMP), subcarrier-to-subcarrier block OMP (BOMP), subcarrier-to-subcarrier pattern-coupled SBL (PCSBL), and polar-domain simultaneous OMP (P-SOMP)~\cite{cui2022channel}. Specifically, P-OMP, BOMP, and PCSBL leverage channel sparsity to perform independent estimation per subcarrier, whereas P-SOMP performs joint (simultaneous) estimation across subcarriers. Note that P-OMP and P-SOMP both employ polar-domain dictionaries.

In the simulations, the BS uses a ULA with 4 antennas, and the XL-RIS adopts a UPA with $N = 128 \times 8$ elements. The carrier frequency is $f_c = 100$ GHz, the bandwidth is $B = 10$ GHz, and the number of subcarriers is $P = 128$. The precoder $\mathbf{F}$ has i.i.d. complex Gaussian entries. The channel components $\mathbf{G}$ and $\mathbf{h}$ are randomly generated. The numbers of BS--RIS and RIS--UE paths are set to $L = 2$ and $K = 2$, respectively, and the Rician factor is 13 dB~\cite{6834753}. For each path, $\{\vartheta,\theta_1,\theta_2,\varphi_1,\varphi_2\}$ are drawn uniformly from $[-\pi/2,\pi/2]$. The RIS is deployed on the user side. The BS--RIS distance is fixed to 50 m to ensure far-field propagation, while the RIS--UE distance is uniformly sampled from $[F_r, R]$, where $F_r = 0.62\sqrt{D^3/\lambda} \approx 0.9433\ \text{m}$ and $R = 2D^2/\lambda \approx 24.267\ \text{m}$, such that the RIS--UE link lies in the near-field region. Performance is evaluated by the normalized mean-squared error (NMSE):
\begin{align}
\text{NMSE (dB)} 
= 10 \log_{10} \left( 
\mathbb{E} \left\{ 
\| \hat{\mathbf{H}} - \mathbf{H} \|_F^2 / \| \mathbf{H} \|_F^2 
\right\} 
\right)
\end{align}

\begin{figure}[!htp]
	\centering
	\subfloat{\includegraphics[width=0.35\textwidth]{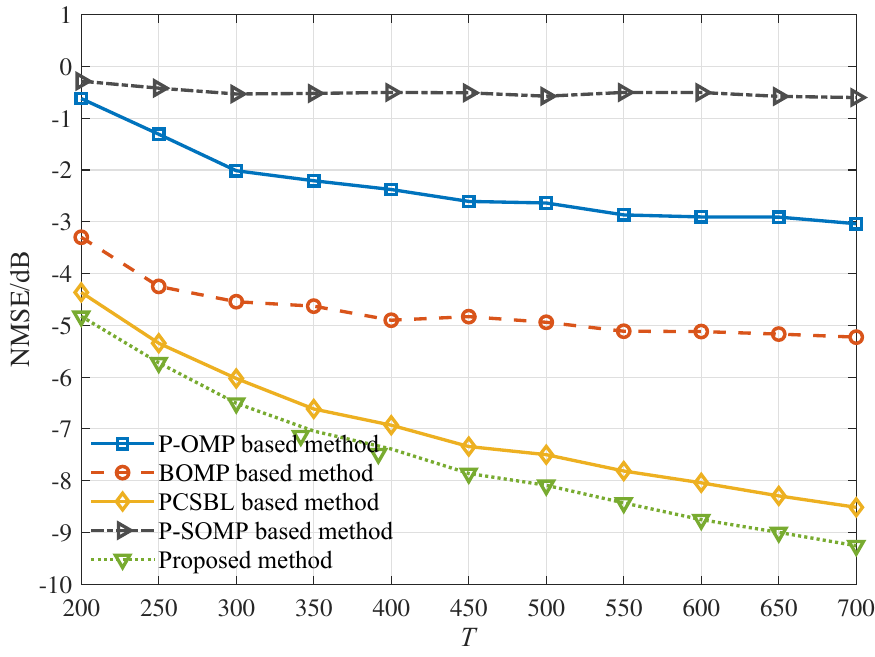}} 
	\hfil
	\caption{NMSE versus pilot length $T$ when SNR = 10 dB.}
	\label{fig:T}
\end{figure}

Fig.~\ref{fig:T} depicts the NMSEs of the competitive algorithms versus the pilot length $T$ when the SNR is set to 10 dB. As $T$ increases, all algorithms exhibit a steady decrease in NMSE. It can be seen that the proposed method surpasses polar-domain methods in estimation accuracy while substantially reducing pilot overhead. It is worth mentioning that the P-OMP approach consistently outperforms P-SOMP, as the wideband effect violates the common-support assumption across subcarriers. Furthermore, compared with the PCSBL variant with single-subcarrier coupling, proposed method lowers NMSE more rapidly, primarily because it introduces a prior that models cross-subcarrier correlations, thereby effectively capturing inter-subcarrier coupling.

\begin{figure}[!htp]
	\centering
	\subfloat{\includegraphics[width=0.35\textwidth]{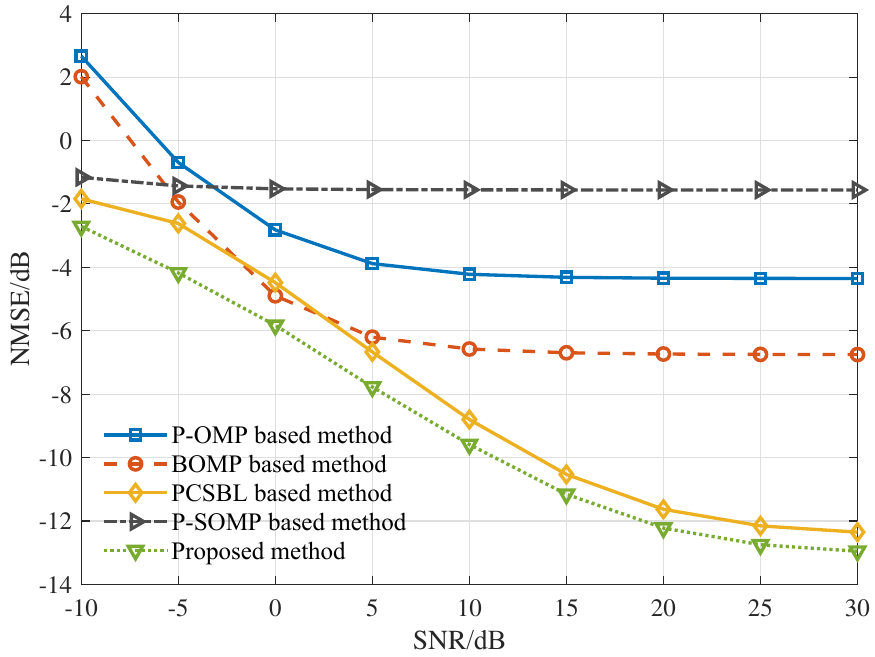}} 
	\hfil
	\caption{NMSE versus SNR when T = 700.}
	\label{fig:SNR}
\end{figure}

Fig.~\ref{fig:SNR} plots the NMSEs versus the SNR when the pilot length $T$ is fixed to $700$. All methods show the improved performance as the SNR increases, but several baselines exhibit clear error floors in the medium-to-high SNR regime. In contrast, the proposed method retains the lowest NMSEs across the entire SNR range, decaying more steeply and leveling off at a lower floor. When SNR is 30 dB, the proposed method significantly outperforms the P-SOMP method with an NMSE gain of approximately 14 dB. This gain stems from the fact that the P-SOMP rigidly enforces a common-support assumption across subcarriers, whereas the proposed method uses a tailored DFT dictionary and explicitly exploits cross-subcarrier correlations, significantly improving near-field wideband channel-estimation accuracy. Collectively, Fig.~\ref{fig:T} and Fig.~\ref{fig:SNR} show that proposed method offers a better accuracy-versus-overhead trade-off than existing polar-domain and sparse-reconstruction schemes, delivering superior overall performance.

\section{Conclusions}
In this work, we investigate the near-field wideband channel estimation problem for XL-RIS assisted communication systems. We adopt a hybrid propagation model, i.e., the spherical-wave model for the XL-RIS–UE link and the planar-wave model for the BS–RIS link, and represent the cascaded channel using Khatri–Rao and Kronecker products. We further construct a frequency-independent orthogonal dictionary to obtain a two-dimensional block-sparse representation across subcarriers, thereby reformulating the problem as compressed sensing with structured sparsity. Numerical results show that the proposed method achieves lower NMSE than existing near-field wideband estimation baselines.

\bibliographystyle{ieeetr}
\bibliography{main} 
\end{document}